\documentclass[12pt,preprint]{aastex}

\newcommand\ie{{\it i.e.\,}}
\newcommand\eg{{\it e.g.\,}}

\newcommand\beq{\begin{equation}}
\newcommand\eeq{\end{equation}}
\newcommand\beqa{\begin{eqnarray}}
\newcommand\eea{\end{eqnarray}}
\newcommand\cc{{\rm cm}^{-3}}

\def\cc{{\rm cm}^{-3}}

\def\Coul{{\rm Coul}}

\def\ln{{\rm ln}}

\begin{document}


\title{Heating and Acceleration of Intracluster Medium Electrons by Turbulence}


\author{Vah\'{e} Petrosian\altaffilmark{1,2} and  William E.
East\altaffilmark{3}}
\affil{Center for Space Science and Astrophysics, Department of Physics,
Stanford University, Stanford, CA 94305}


\altaffiltext{1}{Department of Applied Physics, Stanford University, Stanford,
CA, 94305 email: vahep@stanford.edu}
\altaffiltext{2}{Kavli Institute of Particle Astrophysics and Cosmology,
Stanford University, Stanford, CA, 94305}
\altaffiltext{3}{Department of Physics,
Stanford University, Stanford, CA, 94305 email: weast@stanford.edu}




\begin{abstract}
In this paper we investigate the feasibility of bremsstrahlung radiation from
`nonthermal' electrons as a source of hard X-rays from the intracluster medium
of clusters of galaxies.  With an exact  treatment of the Coulomb collisions in
a Fokker-Planck analysis of the electron distribution we find that the severe
difficulties with lifetimes of `nonthermal' particles  found earlier by Petrosian (2001) using a cold
target model remain problematic. We then address possible acceleration of background electrons into a nonthermal tail. We assume a  simplified but generic
acceleration rate and determine the expected evolution of an initially
Maxwellian distribution of electrons.  We find that strong nonthermal components
arise only for rapid rate of acceleration which also heats up the entire plasma.
These results confirm the conclusion that if the observed `nonthermal' excesses
are due to some  process accelerating the background thermal electrons this
process must be short lived.

\end{abstract}


\section{INTRODUCTION}

The classical picture of the intra-cluster medium (ICM) consisting of nearly
relaxed isothermal hot gas emitting predominantly
the well studied thermal bremsstrahlung (TB) radiation
in the soft
X-ray (SXR,  $\sim 2 - 10$ keV) region has undergone considerable revisions in
recent years. A considerable fraction of the observed clusters appear to be in
the middle of merger process with complex distribution of galaxies. There is
evidence also for considerable deviations
from isothermality; there are hot regions and cold fronts delineated perhaps by
shocks resulting from the merger activity.
An excellent example of this is the cluster RXJ0658, known also as the bullet
cluster, which has achieved considerable notoriety in recent years (see \eg
Markevitch 2005; Brada\^c et al. 2006). In such clusters there is also growing
evidence for nonthermal  activity,
first observed as {\it diffuse radio} radiation
from Coma. Recent systematic searches (see Giovannini et al 1999, 2000) have
detected similar radiation in
more than 40 clusters that are classified either as relic
or halo sources. There is little doubt that this radiation
is due to synchrotron
emission by a population of relativistic electrons.
In the case of
Coma, the radio spectrum may be represented by a broken power law (Rephaeli
1979), or a power law with a rapid steepening (Thierbach et al. 2003) or
with an exponential cutoff
(Schlickeiser et al. 1987) implying the presence of electrons
with similar spectra. Unfortunately,
from radio observations alone one cannot determine the energy of the
electrons or the strength of the magnetic field.  Additional observations or
assumptions are required.  Equipartition or minimum total (particles plus field)
energy arguments imply a population of
relativistic electrons with Lorentz factor $\gamma \sim 10^4$ and magnetic field
strength
of $B\sim \mu{\rm G}$, in rough agreement
with the Faraday rotation measurements (\eg Kim et al. 1990).
Rephaeli and Schlickeiser et al. also pointed out
that these electrons, via inverse
Compton (IC) scattering of the  Cosmic Microwave Background (CMB) photons,
should produce
a broad spectrum of nonthermal  hard
X-ray (HXR) photons (similar to that observed in the radio band) around
50 keV. Detection of HXR radiation could  break the degeneracy
and allow determination of the
magnetic field and the energy of the radiating electrons. In fact, because the
energy density of
the CMB radiation (temperature $T_0$)
$u_{\rm CMB}= 4\times 10^{-13}(T_0/2.8 \, {\rm K})^{4}\,{\rm erg} \,{\rm
cm}^{-3}$
is larger than the  magnetic energy density
$u_{\rm B}= 3\times 10^{-14}(B/\mu{\rm G})^2 \, {\rm erg} \,{\rm cm}^{-3}$,
one expects a higher flux of HXR than radio radiation.

HXR emissions (in the 20 to 80 keV range)
at levels significantly above that expected
from the thermal gas were detected by instruments on board
{\it Beppo}SAX and \textit{RXTE }
satellites from Coma (Rephaeli et al. 1999;
Fusco-Femiano et al. 1999;  Rephaeli \& Gruber 2002;  Fusco-Femiano et al. 2004%
\footnote{The results of this paper have been challenged and rebutted by an
analysis performed with different software by  Rossetti \& Molendi (2004) and
Fusco-Femiano et al. (2007).}),
Abell 2319 (Gruber \& Rephaeli 2002),
Abell 2256 (Fusco-Femiano
et al. 2000; Rephaeli \& Gruber 2003;
and Fusco-Femiano, Landi, \& Orlandini 2005),
and  a marginal ($\sim3\sigma$) detection from Abell 754
and an upper limit on Abell 119 (Fusco-Femiano et al. 2003).
We also note that a possible recent detection of
nonthermal  X-rays, albeit at lower energies, has been reported from a
\underline{poor} cluster, IC 1262, by Hudson et al. (2003).  All
these clusters are nearby clusters in the redshift range
$0.023<z<0.056$.
Notable recent exceptions at higher redshifts
are {\it RXTE} observations of RXJ0568 ($z=0.296$, Petrosian et al. 2006;
PML06) and Abell 2163 ($z=0.208$, Rephaeli, Gruber, \& Arieli 2006)
where the HXR flux is consistent with the
upper limit set by {\it Beppo}SAX (Feretti
et al. 2001)%
\footnote{Note that
the high redshift observations are made relatively easier because of the (known)
rapid increase with redshift of the CMB density (see PML06), so that, in
principle,
the cosmological evolution of these
quantities can be investigated with simultaneous radio and HXR observations.}.

It should also be noted that excess radiation was
detected in the 0.1 to 0.4 keV band by {\it Rosat, Beppo}SAX and
{\it XMM-Newton} and in the EUV region (0.07 to 0.14 keV) and similar
excess radiation was detected by the {\it Extreme
Ultraviolet Explorer} from Coma (Lieu et al.  1996) and some other clusters.
A cooler ($kT\sim 2$ keV) component or IC scattering of
CMB photons by lower energy ($\gamma \sim 10^3$)
electrons are two possible ways of producing this excess radiation.
However, some of the observations and the emission process
are still controversial (see Bowyer 2003).

Even though the IC interpretation seems natural, there is some difficulty with
it.
Soon after the discovery of HXRs it was realized that the relatively high
observed fluxes
require large numbers of relativistic electrons, and consequently a relatively
low  magnetic field for a given observed radio flux.
For Coma, this requires a (volume averaged) magnetic field of
${\bar B}\sim 0.1 \, \mu$G,
while equipartition gives ${\bar B}\sim 0.4 \, \mu$G
and Faraday rotation measurements give the (average line-of-sight) field of
${\bar B}_{\rm l}\sim 3 \, \mu$G
(Giovannini et al.  1993, Kim et al. 1990, Clarke et al.  2001;  2003).
(In general the Faraday rotation measurements of most clusters
give $B> \mu$G; see \eg Govoni et al. 2003.)
Consequently,  various authors
(see \eg En\ss lin, Lieu, \& Biermann 1999; Blasi 2000) suggested
that the HXR radiation is due to nonthermal  bremsstrahlung (NTB) by a
second population of nonthermal  electrons with a power law distribution in the
10 to 100 keV range. However, as shown by Petrosian (2001)
(P01 for short), this process faces a more serious difficulty, which is  hard to
circumvent. This is
because bremsstrahlung is a very inefficient process.
Compared to Coulomb losses the bremsstrahlung yield is very small. For a
particle of energy $E$ much larger than that of background particles $Y_{\rm
brem}
\sim
3\times10^{-6}(E/25 \, {\rm keV})^{3/2}$ (see Petrosian 1973).
Thus, for continuing
production of a HXR
luminosity of $4\times 10^{43}$ erg s$^{-1}$ (observed for Coma),
a power of $L_{\rm
HXR}/Y_{\rm brem} \sim 10^{49}$ erg s$^{-1}$ must be continuously fed into
the ICM, increasing its
temperature to $T\sim 10^8$ K after $3\times 10^7$ yr, or to $10^{10}$ K in a
Hubble time indicating that the NTB emission phase must be very short
lived. As pointed out in P01, a corollary of this is that it would be difficult
to accelerate thermal particles to produce a nonthermal tail without excessive
heating of the background plasma.

The above arguments, however, are not definitive.

  1. The argument against the IC model is not
as severe as stated above. There are several factors which may
resolve the apparent discrepancy among different estimates of the magnetic
field. Firstly, the $B$ field value based on the Faraday rotation measure
assumes a chaotic magnetic
field with scale of few kpc which is not a directly measured quantity
(see \eg Carilli \& Taylor 2002).  Secondly, the accuracy of the quoted
measurements
have been questioned by Rudnick \& Blundell (2003) and defended by
Govoni \& Feretti (2004) and others. Thirdly, as pointed by
Brunetti et al. (2001), a strong gradient in
the magnetic field can reconcile the difference between the volume and
line-of-sight averaged measurements.
Finally, as pointed out in P01, this discrepancy can be alleviated by
a more realistic electron spectral distribution (\eg the spectrum
with exponential cutoff suggested by Schlickeiser et al. 1987) and/or
a non-isotropic pitch angle
distribution.   In addition, for a population
of clusters  observational selection
effects come into play and may favor Faraday rotation detection
in high $B$ clusters which will have
a weaker IC flux relative to synchrotron.

2. The spectral shape of the HXR emission is not very well
constrained so that a two temperature model fits the observation as well as a
single temperature plus a power law model (see \eg PML06). However, the second
thermal component
of electrons must have a much higher temperature
than the gas responsible for the SXR emission.
For production of HXR flux up to 50 keV
this requires a gas with $kT>30$ keV  and
(for Coma) an emission measure about 10\% of
that of the SXR producing plasma. Heating and
maintaining of the plasma to such high
temperatures in view of rapid equilibration expected
by classical Spitzer conduction suffers from the same shortcoming as the NTB
case%
\footnote{A possible way to circumvent the rapid cooling of the hotter plasma by
conduction or rapid energy loss of the nonthermal  particles is to
physically separate
these from the cooler ICM gas. Exactly how this can be done
is difficult to determine but strong
magnetic fields or turbulence may be able to produce such a situation.}.
In fact, as we shall see below, the thermal and nonthermal  scenarios cannot be
easily distinguished from each other. The acceleration mechanism energizes the
plasma and modifies its distribution in such a way that both heating and
acceleration
take place.

3. The short timescale estimated above  is based on energy losses of electrons
in a cold
plasma  which is a good approximation for electron energies
$E\gg kT$. As $E$ nears $kT$ the rate of Coulomb loss (mainly due to
electron-electron collision) decreases but the
bremsstrahlung rate (due to the electron-proton collision at these
nonrelativistic energies) remains constant. There has been several attempts to
address this issue. Blasi (2000) using a more realistic treatment of the Coulomb
collison in a Fokker-Planck treatment, based on coefficients derived by
Nayakshin \& Melia (1998) (NM98, for short),  produced a nonthermal tail in the
electron distribution which might explain the HXR observations from the Coma
Cluster. Wolfe \& Melia (2006), on the other hand, expanding on the results from
NM98 use a covariant treatment of the kinetic equation find that the the result
of energizing of the plasma by turbulence is primarily to heat the plasma to
higher temperatures on a short timescale in agreement with P01.
Finally, in a recent paper Dogiel et al. (2007) claim
that
in spite of the short  lifetime of the test particles due to their Coulomb loses
the `particle
distribution'
lifetime is longer and a power law tail can be maintained without requiring the
energy input estimated above.

In this paper we address this problem not with
the test particle and cold plasma assumption but by carrying out a realistic
acceleration and energizing calculation of the ICM plasma by turbulence or any
similar mechanism.
In \S 2 we describe our method of evaluating the influence of turbulence (or any
other acceleration process) and Coulomb collisions on the spectral distribution
of electrons in a hot
plasma appropriate for the ICM. In \S 3 we first present a test of our algorithm
and then address the question of the lifetime of nonthermal tails. In \S 4 we
apply the method to the acceleration of thermal background particles by a
generic acceleration model
and present some results on the evolution of the
distribution of electrons and estimate the fraction of electrons that can be
considered as nonthermal. In \S 5 and  summarize our results, compare them with
those from previous works and present our
conclusions.
In the Appendix we describe some technical details of our procedure.

\section{BASIC SCENARIO OF ACCELERATION}

In this section we consider a hot gas subject to some acceleration process. In a
cluster the hot gas is confined by the gravitational field of the total  (dark
and `visible') matter. Relativistic particles, on the other hand, can cross the
cluster of radius $R$ on a timescale of $T_{\rm cross} = 3\times 10^6(R/$Mpc) yr
and can escape the cluster unless confined by a chaotic magnetic field or a
scattering agent, such as turbulence, with a mean free path $\lambda_{\rm scat}
\ll R$. For confinement on a Hubble timescale of $10^{10}$ yr we need
$\lambda_{\rm
scat} < 10$ kpc. As stated above the magnetic field is expected to be  chaotic
and there are good arguments for the presence of turbulence, especially in
clusters
with recent merging episodes. For example XMM-Newton observations indicate that in the Coma cluster more than $10\%$ of the ICM pressure is in turbulent form (Schuecker et al. 2004).  In addition, modeling of the circular motion of the galaxies within clusters has also indicated that this can give rise to turbulence (see \eg  Kim 2007).

As a result of scattering from this turbulence the particle
pitch angle changes stochastically with the diffusion rate $D_{\mu\mu}$  ($\mu$
stands for the cosine of the pitch angle). When the scattering time
$\tau_{\rm scat}=\lambda_{\rm scat}/v\sim \langle 1/D_{\mu\mu}\rangle$ is much
less than the
dynamic and other timescales of the particles (with velocity $v=c\beta$), the
pitch angle distribution of the particles will be isotropic. Also as a result of
this scattering, particles will be accelerated stochastically on a time scale of
$\tau_{\rm diff}=p^2/D_{pp}$,
where $D_{pp}$ is the
momentum diffusion coefficient. Particles may also undergo direct acceleration
at a rate of say $A(E)$ or timescale $\tau_{\rm ac}\sim E/A(E)$ and will lose
energy
at a rate of $\dot{E}_{L}$ due to their interactions with background particles
and fields%
\footnote{In what follows, except when specified, all energies will be expressed
in units of electron rest mass energy $m_ec^2$ so that $E=\gamma-1$. Note that
this also means that $A(E)$, the direct energy gain rate (as well as the loss
rates discussed below), will be in units of $m_ec^2/$s, and the diffusion
coefficients in units of $(m_ec^2)^2/$s.}.

The evolution of the energy and pitch angle distribution function of plasma
particles subjected to a stochastic acceleration process integrated over the
volume of the turbulent region can then be described by the
Fokker-Planck transport equation.
Under the conditions specified above this equation simplifies considerably. The
transport equation describing the
gyrophase and pitch angle averaged spectrum, $N(E,t)$, of the particles can be
written as (see \eg Petrosian \& Liu 2004)%
\begin{equation} \label{KEQ}
\frac{\partial N}{\partial t} = \frac{\partial ^2}{\partial E^2} [(D(E)+D_{\rm
Coul}(E)) N] -
\frac{\partial}{\partial E}[(A(E)  - \dot{E}_{L})N].
\end{equation}
For stochastic acceleration by turbulence%
\begin{equation} \label{coeff}
D(E)=\beta^2 D_{pp}\,\,\,\,\, {\rm and}\,\,\,\,\, A(E)=D(E)\zeta(E)/E+dD(E)/dE
\end{equation}
describe the diffusive
and systematic accelerations coefficients.
Their value and evolution is
determined by the energy density and spectrum of turbulence. Here
$\zeta(E)=(2-\gamma^{-2})/(1+\gamma^{-1})$ is a slowly varying
function
changing from 1/2 to 2 for $0<E<\infty$. The term  $dD(E)/dE$ would be
absent if the diffusion term in
equation (\ref{KEQ}) were written as $\frac{\partial }{\partial E} [D(E)
\frac{\partial
}{\partial E} N(E)]$ which is another commonly used form of the transport
equation. The numerical results presented below are based on the code developed
by  Park \& Petrosian (1995, 1996) that uses this
form of the equation.

\begin{figure}[htbp]
\leavevmode\centering
\includegraphics[height=7cm]{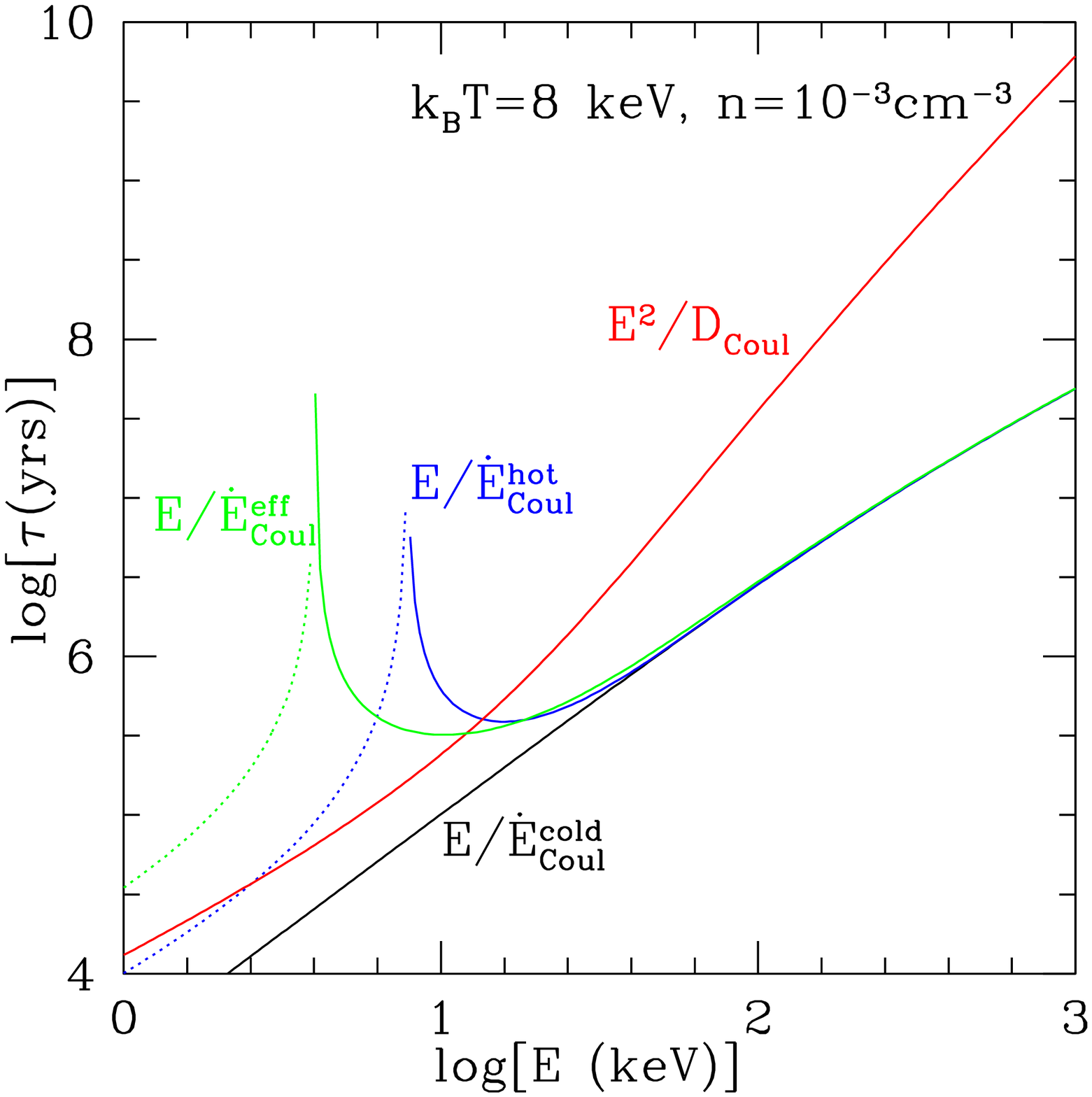}
\includegraphics[height=7cm]{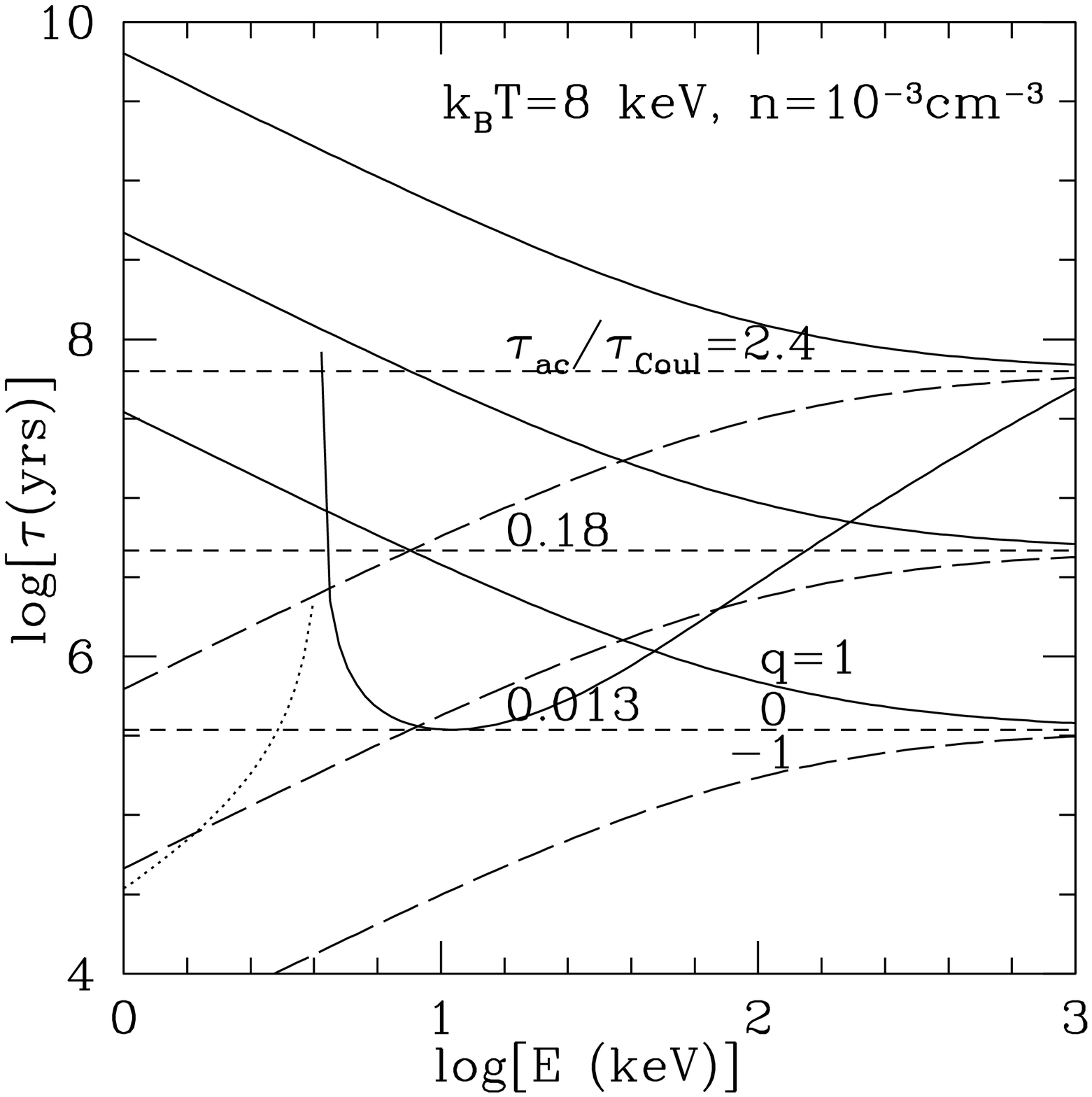}
\caption{{\bf Left panel:} Various timescale for Coulomb
collisions for a hot plasma with typical ICM parameters from equations
(\ref{cold}), (\ref{hot}), (\ref{DCoul}) and (\ref{eff}), from Liu (2006). {\bf
Right panel:}
Acceleration timescale based on the model described by equation (\ref{acctime})
for $E_c=0.2$ and the specified parameters, and the total loss timescale for ICM
conditions. We
use the effective Coulomb loss rate given
in equation (\ref{eff}). For completeness the IC plus synchrotron losses for a
CMB temperature of $T_{\rm CMB}=3$ K and ICM magnetic field of $B=1\mu$G are
also included but their influence appears at $E>10^4$ keV (see \eg  P01).}
\label{timescales}
\end{figure}

In what follows we will not be concerned with the exact forms of
these coefficients and will assume some very simple energy dependence. We will
assume them to be constant in time which is equivalent to having a constant
density and spectrum of turbulence. Specifically we will assume
\begin{equation} \label{diffcoeff}
D(E)=\frac{E^2}{\zeta(E)\tau_0(1+E_c/E)^q}
\end{equation}
so that  for the alternate form of the transport equation used in our numerical
code we have a simple acceleration time
\begin{equation}\label{acctime}
\tau_ {ac}=\tau_0(1+E_c/E)^q.
\end{equation}
The right panel of
Figure
\ref{timescales} shows the acceleration time with $E_c=0.2 (\sim 25$ keV) for
several values of the
parameters $\tau_0$ and $q$ appropriate for ICM condition along with the Coulomb loss rate discussed in the next section.  The rate at which energy would be added to the particles due to such turbulence is given by $\int_0^{\infty}A(E)N(E,t)dE$ (see appendix) which, assuming a particle distribution normalized to one, is approximately $\langle E\rangle/\tau_0$ where $\langle E \rangle$ is the average energy.
\clearpage
\clearpage

The remaining coefficient  $\dot{E}_{L}$ is sum of the loss rates (defined here
to be positive) due to Coulomb collisions (primarily with background electrons),
synchrotron, IC scattering (of CMB photons) and bremsstrahlung. We will include
all these terms in our analysis but for nonrelativistic energies in the ICM the
dominant
term is due to Coulomb collisions, and at low (mainly nonrelativisitic)
energies, which will be our main focus here, the Coulomb term will be the most
important one (see Figure \ref{timescales}). As mentioned above, the
previous analysis was based on energy loss rate due to Coulomb collisions
with a ``cold" ambient plasma (target electrons having zero velocity):
\begin{equation}
\dot{E}_{\rm Coul}^{\rm cold} = 1/(\tau_{\rm Coul}\beta), \,\,\,\,\,\, {\rm
where}\,\,\,\,\,\, \tau_{\rm Coul}\equiv (4 \pi r_0^2 \ln{\Lambda} c n)^{-1},
\label{cold}
\end{equation}
and $r_0=e^2/(m_e c^2)=2.8\times10^{-13}$ cm is the classical electron
radius.  For ICM  conditions the Coulomb logarithm $\ln{\Lambda}\sim 40$ and
density $n\sim 10^{-3}$ cm$^{-3}$, therefore $\tau_{\rm Coul}\sim 2.7\times
10^7$ yr. The cold target loss rate is a good approximation when the nonthermal
electron velocity $v \gg v_{\rm th}$,
where $v_{\rm th}= \sqrt{2kT/m_e}$
is the thermal velocity of the background electrons.  This approximation becomes
worse as  $v \rightarrow v_{\rm th}$ and breaks down completely for $v <
v_{\rm th}$,
in which case, the electron may  gain energy
rather than lose energy as is always the case in the cold-target scenario.
More general treatment of Coulomb loss is therefore desired. For a hot plasma
the above loss rate must be modified and there will also be a non-zero Coulomb
diffusion
term $D_{\rm Coul}(E)$.

Let us first consider the {\it energy loss rate}. This is obtained from the rate
of exchange of energy between two electrons with energies $E$ and $E'$ which we
write as
\begin{equation}
\Delta E/\Delta t =  G(E,E')/\tau_{\rm Coul}.
\label{exchange}
\end{equation}
Here $G$ is an antisymmetric function of the two variables such that the higher
energy electron loses energy and the lower energy one gains energy. From
equations (24)-(26) of NM98 we can write
\begin{equation} \label{cases}
G(E,E')=\cases{-\beta'^{-1},&if $E'>E, E\ll 1$;\cr
                \beta^{-1},&if $E'<E, E'\ll 1$;\cr
		E'^{-1}-E^{-1},&if $E, E'\gg 1$.}
\end{equation}
The general Coulomb loss term is obtained by integrating over the particle
distribution:
\begin{equation}\label{general}
\dot{E}_{\rm Coul}^{\rm gen}(E,t) ={1 \over \tau_{\rm Coul}}\int_0^\infty
G(E,E')N(E',t)dE'.
\end{equation}
Similarly, we can express the {\it Coulomb diffusion coefficient} as
\begin{equation}\label{generaldiff}
D_{\rm Coul}^{\rm gen}(E,t) ={1 \over \tau_{\rm Coul}}\int_0^\infty
H(E,E')N(E',t)dE'.
\end{equation}
From equations (35) and (36) of NM98%
\footnote{Note that the first term in equation (35) should have a minus sign and
that the whole quantity is too large by a factor of 2; see also
Blasi 2000 for other typos.}
we get
\begin{equation} \label{casesdiff}
H(E,E')=\cases{\beta^2/(3\beta'),&if $E'>E, E\ll 1$;\cr
               \beta'^2/(3\beta),&if $E'<E, E'\ll 1$;\cr
		1/2,&if $E, E'\gg 1$.}
\end{equation}
Thus, the determination of the distribution $N(E,t)$ involves solution of the
combined integro-differential equations (\ref{KEQ}), (\ref{general}) and
(\ref{generaldiff}), which
can be solved iteratively. However, in many cases these equation can be
simplified considerably.
For example, if we are interested only in the ``supra-thermal" tail of the
distribution, where the energy $E$ is larger than that of the bulk of the
population, and if these are mainly nonrelativisitic, $\langle E' \rangle \ll
1$, then we can use the approximation $G(E,E')=\beta^{-1}$ and
$H(E,E')=2E'/\beta\ll G(E,E')$ (the second lines in
eq. [\ref{cases}] and  [\ref{casesdiff}]). In this case the Coulomb loss term is
given by equation
(\ref{cold}), the Coulomb diffusion term is absent and we have a simple
differential equation to solve. Another
simplification arises  when the bulk of the particles have Maxwellian
distribution $N(E')=n(2/\sqrt{\pi})(kT/m_ec^2)^{-3/2}E'^{1/2}e^{-E'm_ec^2/kT}$,
with $kT\ll m_ec^2$.
Carrying out the integration of equations (\ref{general}) and
(\ref{generaldiff}) over this  energy distribution, and after some algebra, the
net energy loss (gain) and diffusion coefficient can be written as (see also
Spitzer 1962,
pp.128-129; Benz 2002, eq. 2.6.28; Miller et al. 1996)
\begin{equation}
\dot{E}_{\rm Coul}^{\rm hot} = \dot{E}_{\rm Coul}^{\rm cold}
     \left [{\rm erf}(\sqrt{x}) - 4 \sqrt{{x \over \pi}} e^{-x} \right ],
\label{hot}
\end{equation}
and
\begin{equation}
 D_{\rm Coul}(E) = \dot{E}_{\rm Coul}^{\rm cold} \left({kT \over m_e c^2}\right)
    \left[ {\rm erf}(\sqrt{x}) - 2 \sqrt{{x \over \pi}} e^{-x} \right],
\,\,\,\,\,\,{\rm with} \,\,\,\,\,\, x\equiv {Em_ec^2 \over kT},
\label{DCoul}
\end{equation}
where ${\rm erf}(y) = {2 \over \sqrt \pi} \int_0^y e^{-t^2} dt$ is the error
function.

We should note that for the form of the kinetic equation that we use for our
numerical
results we should include a term equal to $(dD/dE)_{\rm Coul}$  in the
second right hand term of eq.(\ref{KEQ}). This is equivalent to defining an
effective loss rate for the code
\begin{equation} \label{eff}
{\dot E}_{\rm Coul}^{\rm eff} = {\dot E}_{\rm Coul}^{\rm hot} + {d D_{\rm Coul}
\over d E} = \dot{E}_{\rm Coul}^{\rm cold} \left[ {\rm erf}(\sqrt{x}) - 2
\sqrt{{x \over \pi}} e^{-x} \right]\left( 1 - \frac{1}{\gamma(\gamma
+1)x}\right),
\end{equation}
where we have used equations (\ref{hot}) and (\ref{DCoul}). Note that for
relativistic test particles we have the same expressions
as long as $kT\ll m_ec^2$.
The various Coulomb rates above are shown for typical ICM conditons on the left
panel and the effective Coulomb loss rate (the sharply peaked curve) is shown along with the acceleration timescale in the right panel of Figure
\ref{timescales}. Here solid lines show energy loss and dotted lines are for
energy gain.

\section{RELAXATION AND THERMALIZATION TIMESCALE}

\begin{figure}[htbp]
\begin{center}
\includegraphics[height=7cm]{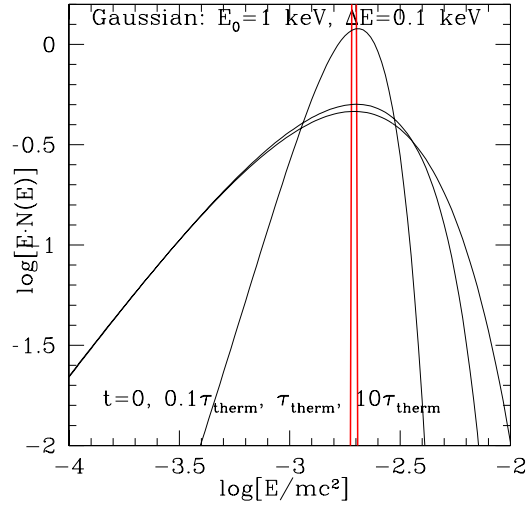}
\end{center}
\caption{ Evolution of an initial narrow Gaussian distribution of
electrons (heavy red line) with $E_0=1$ keV subject to elastic Coulomb
collisions (loss and
diffusion at a rate calculated using the final expected temperature $kT=2E_0/3$)
showing approach
to a Maxwellian distribution within several thermalization
timescales $\tau_{\rm therm}\simeq 2\tau_{\rm Coul}E_0^{1.5}$.}
\label{thermalization2}
\end{figure}

As a test of our
algorithm we first address the relaxation into a thermal distribution
of particles with an initial Gaussian  distribution  (mean energy $E_0$ and
width $\Delta E \ll E_0$) subject only to inelastic Coulomb collisions. The
distribution should approach a Maxwellian with $kT/m_ec^2=2E_0/3$ and total
number and energy equal to that of the initial particles after several
thermalization times (Spitzer 1962, Benz 2002)
\beq\label{thermalization}
\tau_{\rm therm}=3.5\tau_{\rm Coul}(kT/m_ec^2)^{1.5}.
\eeq
Using only the Coulomb loss and diffusion terms for this value of the
temperature, in Figure \ref{thermalization2} we
show the evolution of a initial narrow Gaussian electron spectrum toward the expected Maxwellian
distribution. As evident, most of the particle settle down into a thermal
distribution within
several thermalization times. This is similar to the result found by Miller et
al.
(1996).

\clearpage
\clearpage

\begin{figure}[htbp]
\begin{center}
\includegraphics[height=7cm]{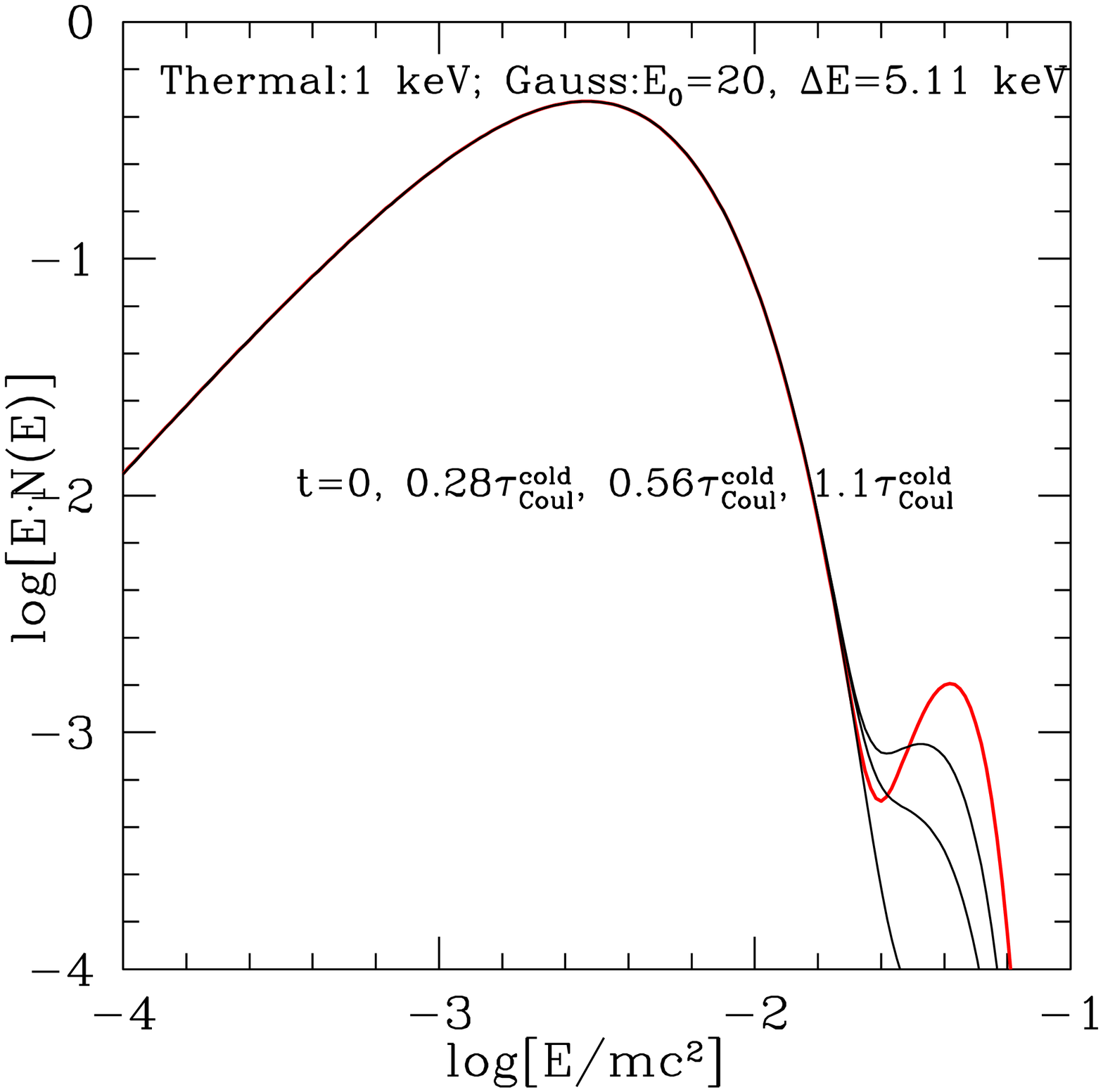}
\includegraphics[height=7cm]{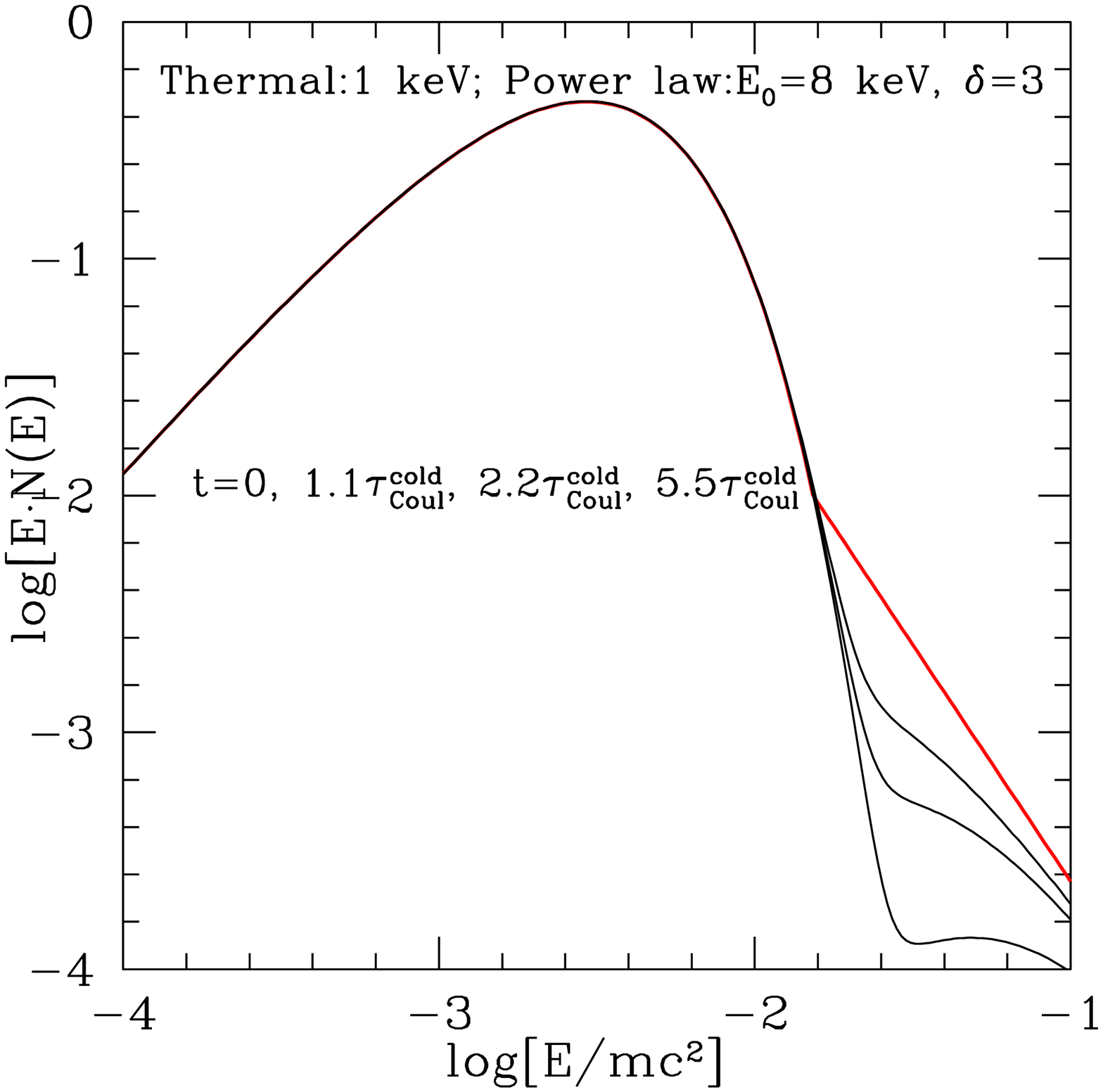}
\includegraphics[height=7cm]{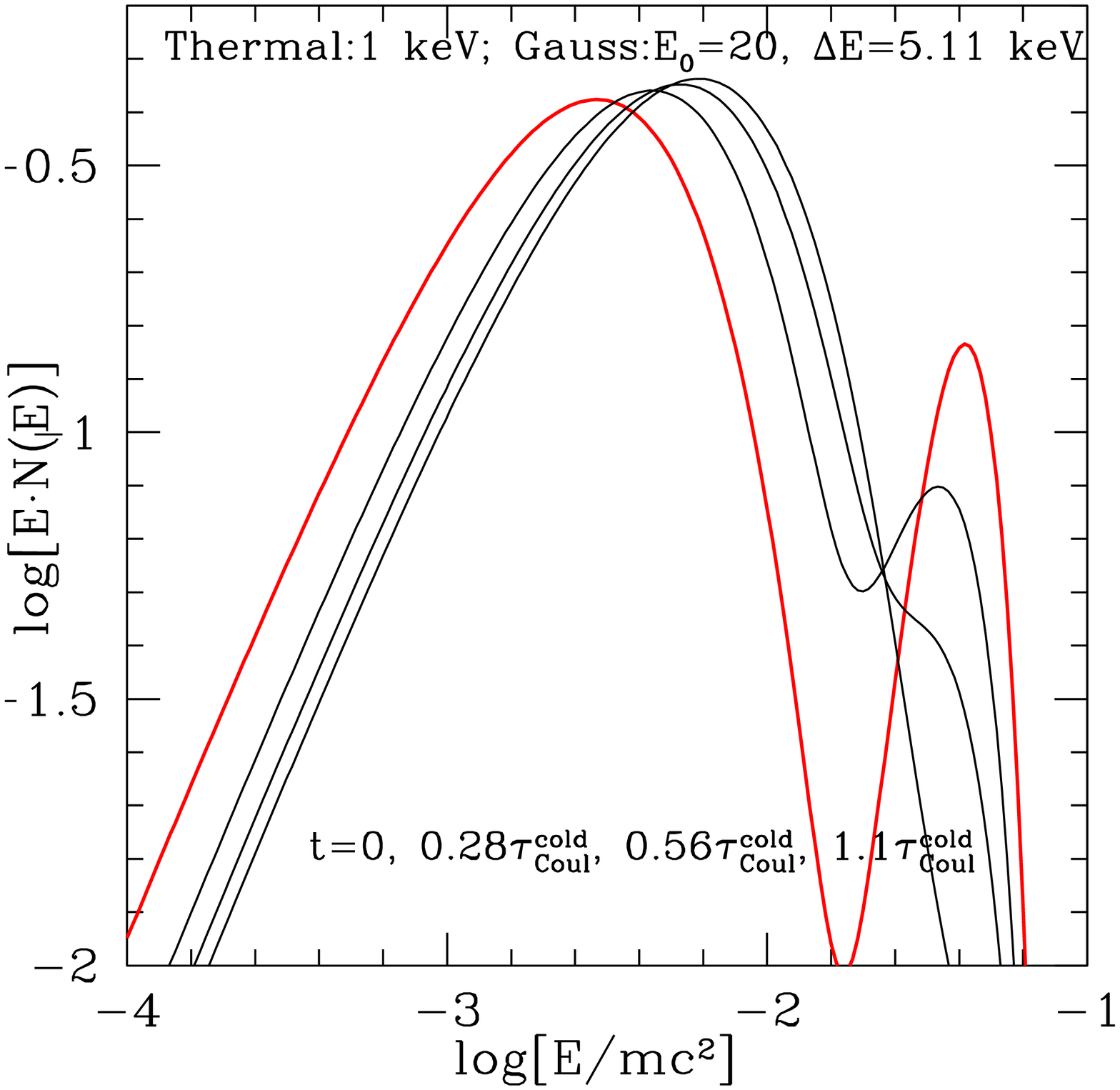}
\includegraphics[height=7cm]{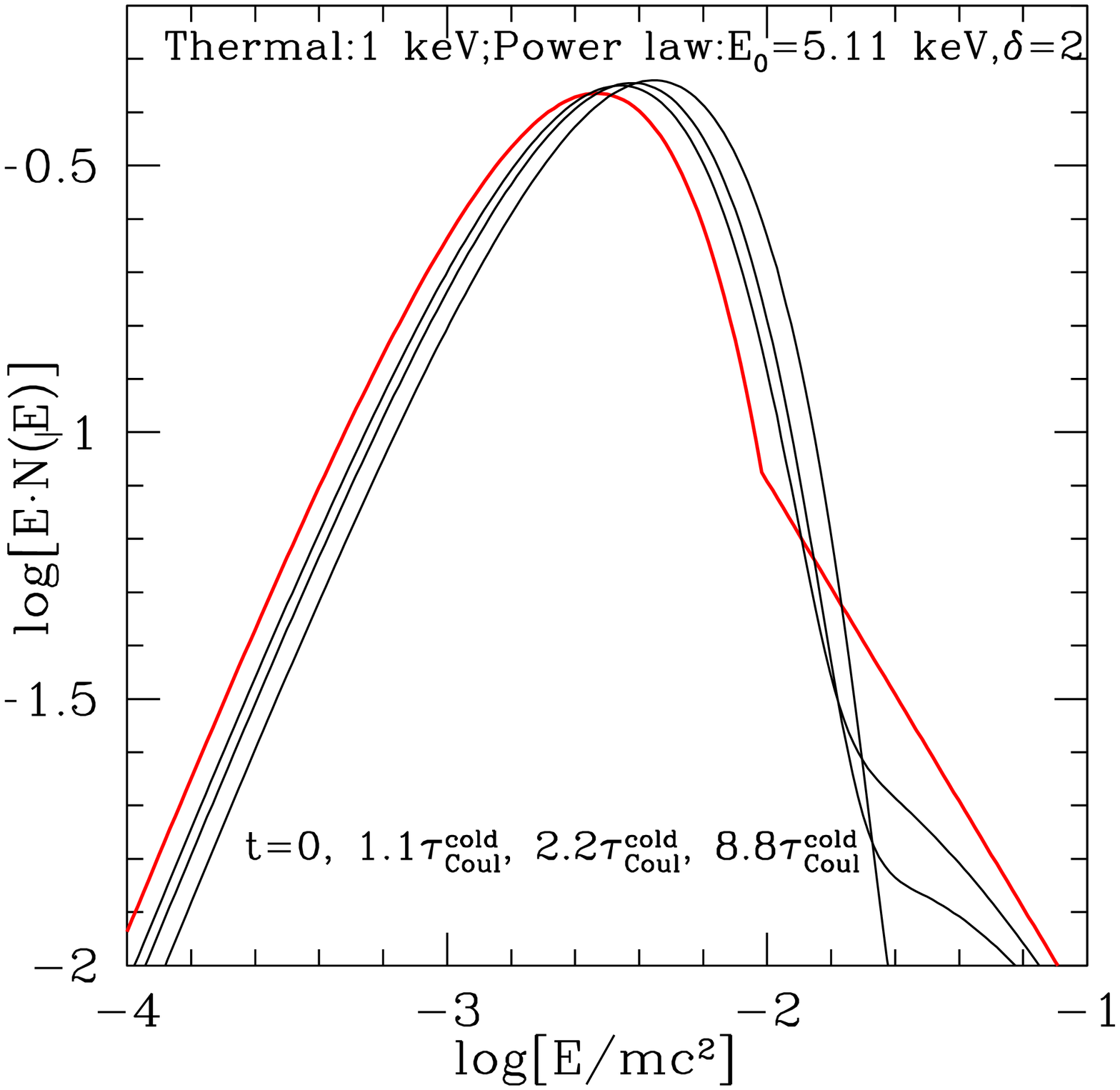}
\end{center}
\caption{ {\bf Top panels:} Evolution of an initial nonthermal and
isotropic
distribution of electrons (heavy red line) subject to elastic Coulomb collisions
primarily with a
background thermal electrons with temperature $kT=1$ keV showing gradual
degradation of the nonthermal tails. {\bf Left panel} for a narrow Gaussian
distribution with mean energy $E_0=20$ keV and {\bf right panel} for a power law
tail
with spectral index $\delta=3$ for $E_0\geq 8$ keV. In both cases the nonthermal
tails are reduced by
a factor of ten within
less than three times the cold target coulomb loss time
$\tau^{\rm cold}_\Coul\equiv E/{\dot E}^{\rm cold}_\Coul=\sqrt{2}\tau_\Coul
E_0^{1.5}$ for
appropriate energy $E_0\sim 20$keV. {\bf Bottom panels:} Same as above but for
nonthermal particles
having a significant energy so that their thermalization heats the plasma to a
higher temperature. The  thermalization time is same as the `relaxation' above.}
\label{tests}
\end{figure}

Next we consider {\it thermalization }or {\it energy loss timescale} of
nonthermal populations of electrons
(with isotropic pitch angle distribution) added to a  background
thermal plasma. We first  consider  what one may call the test particle
case where the nonthermal tails contain a much smaller amount of energy than the
background particles. Or alternatively we assume that the energy lost by the
injected nonthermal tail is radiated or conducted away so that the background
temperature
stays constant. In this case we can use the thermal form of the coefficients
given by equations (\ref{DCoul}) and (\ref{eff}) calculated for the constant
temperature.  We consider two different forms of  nonthermal tails; one a {\it
Gaussian spectrum} of electrons%
\footnote{For the purpose of this test we ignore the fact that such a
bump-on-tail distribution is unstable and will give rise to plasma turbulence
(Langmuir waves) which will modify the tail into a plateau within a short
timescale $\sim n/(\Omega_pn_{\rm bump}$), where $n_{\rm bump}$ is the density in
the tail and $\omega_p=\sqrt{2\pi ne^2/m}\sim 2\times 10^3$ Hz for typical ICM
density $n\sim 10^{-3}$ cm$^-3$.}
with mean energy $E_0$ and width $\Delta E$ and another a power-law tail
starting at some energy $E_0>kT$.  The top two panels of Figure \ref{tests}
show the evolutions of these nonthermal tails. As
evident, the nonthermal
distribution is depleted to a tenth of its original size within several
cold target Coulomb loss  times,
$\tau^{\rm cold}_\Coul\equiv E/{\dot E}^{\rm cold}_\Coul=\sqrt{2}\tau_{\rm
Coul}E_0^{1.5}$, appropriate for an initial energy $E_0$.

\clearpage
\clearpage

Finally we consider similar cases to those above but without the ``test
particle"
assumption. In these cases the nonthermal
tails contain a significant amount of energy so that the energy lost by them
heats the
plasma and changes its temperature. This in turn changes the loss and
diffusion coefficients. We evaluate the total particle evolution using two
different methods.
In the first method at each time step we obtain an estimated  temperature and
use the new temperature to update the coefficients according to equations
(\ref{DCoul}) and (\ref{eff}).  Our procedure for advancing the temperature  to
its new  value is described in the
appendix. The second and more accurate but also more time consuming method is to
use equations (\ref{general}) and (\ref{generaldiff}) to calculate the values of
the coefficients at each time step. The results obtained from these two methods
are essentially indistinguishable (see appendix for comparison).
The bottom panels of Figure \ref{tests} show the evolution of the total (thermal
plus
tail) spectrum of electrons with initial forms similar to those shown in Figure
\ref{tests}. As the tail is dissipated, the temperature increases to its
final value within a time of  $100\tau_{\rm therm}$ of the original temperature
of 1 keV or 2 to 3 times the cold target loss time for $E_0=20$ keV particle.

The above results show that the conclusions based on the cold plasma
approximation are good order of magnitude estimates and that using the more
realistic hot plasma relations changes these estimates by factors of two or
three. Consequently, the estimates made in P01 based on cold target assumption
are
modified by similar factors; the required input energy will be lower
and the time scale for heating will be longer by the same factor.
This agrees qualitatively with Figure 3 of Dogiel et al. (2007) but does not
support
their other claims about long lifetimes of power-law tails which are based on a
less realistic treatment of the problem (see below).

\section{HEATING AND ACCELERATION OF ELECTRONS}

In this section we investigate the evolution of spectra of ICM electrons
subject to diffusion and acceleration by turbulence and diffusion  and energy
losses due to Coulomb collisions using the equations described in \S 2. We also
include synchrotron, IC and bremsstrahlung losses which have insignificant
effect on the final results for the ICM conditions. We start with a ICM of
$kT=8$ keV and $n=10^{-3} \cc$ and assume a continuous injection of turbulence
with
a rate so
that its energy density remains constant resulting in time independent diffusion
and
acceleration rates (\ie parameters $q$, $E_c$, and $\tau_0$ in eq.
[\ref{acctime}] are
constants). However, the Coulomb coefficients  must be updated. Again we use
two different approaches. In the first, at each time step we estimate a new
temperature  using the fitting prescription described in the second part of  the
appendix and calculate the coefficients based on equations (\ref{DCoul}) and
(\ref{eff}). This is accurate at low acceleration rates where the deviation
from a Maxwellian distribution is small. But at higher rates these deviations
become large and we use the more accurate method described above. At each time
step we use the new
distribution of the electrons and equations (\ref{general}) and
(\ref{generaldiff}) to calculate the values of the Coulomb coefficients.

\begin{figure}[htbp]
\begin{center}
\includegraphics[height=6.2cm]{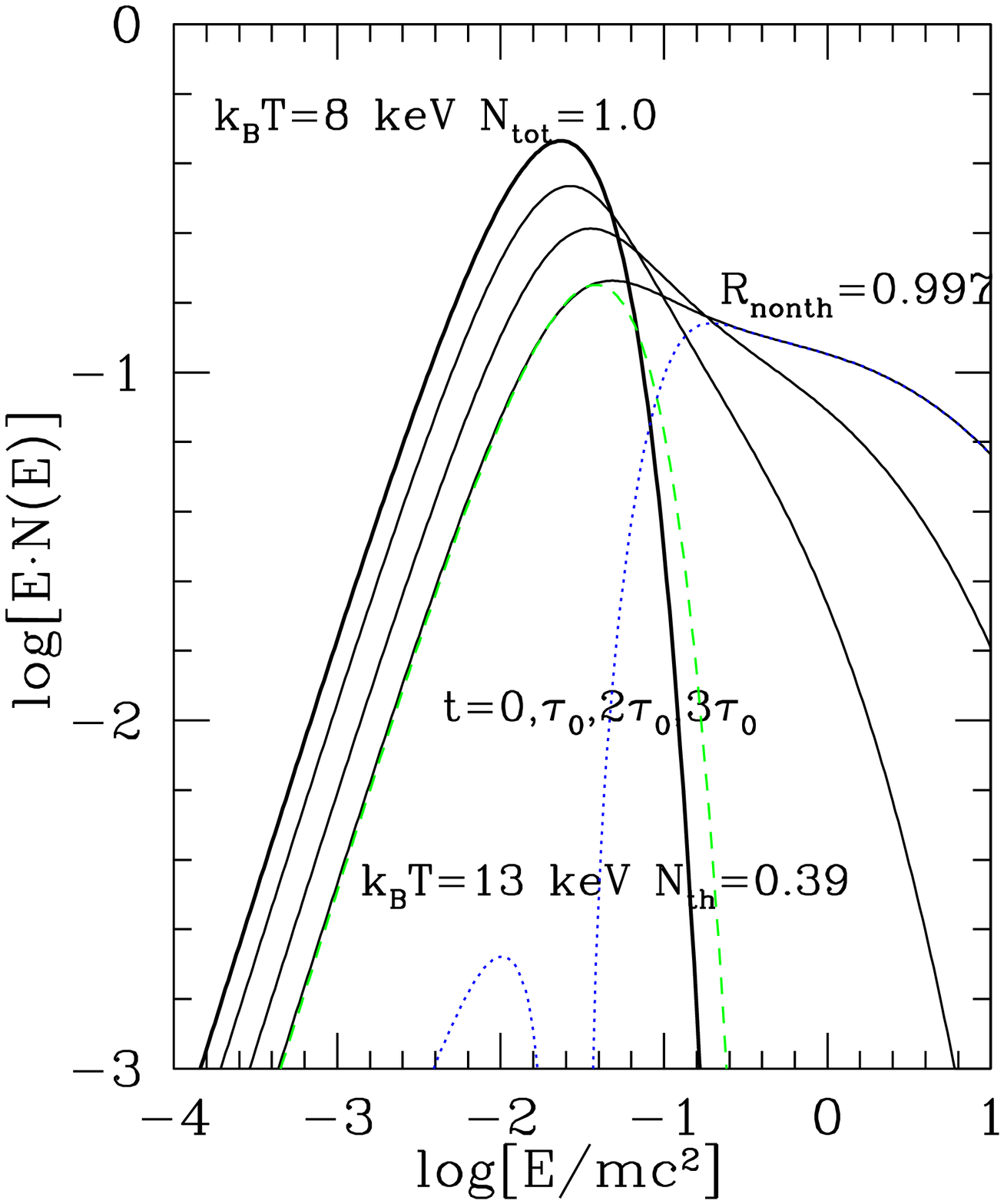}
\hspace{-1.4cm}
\includegraphics[height=6.2cm]{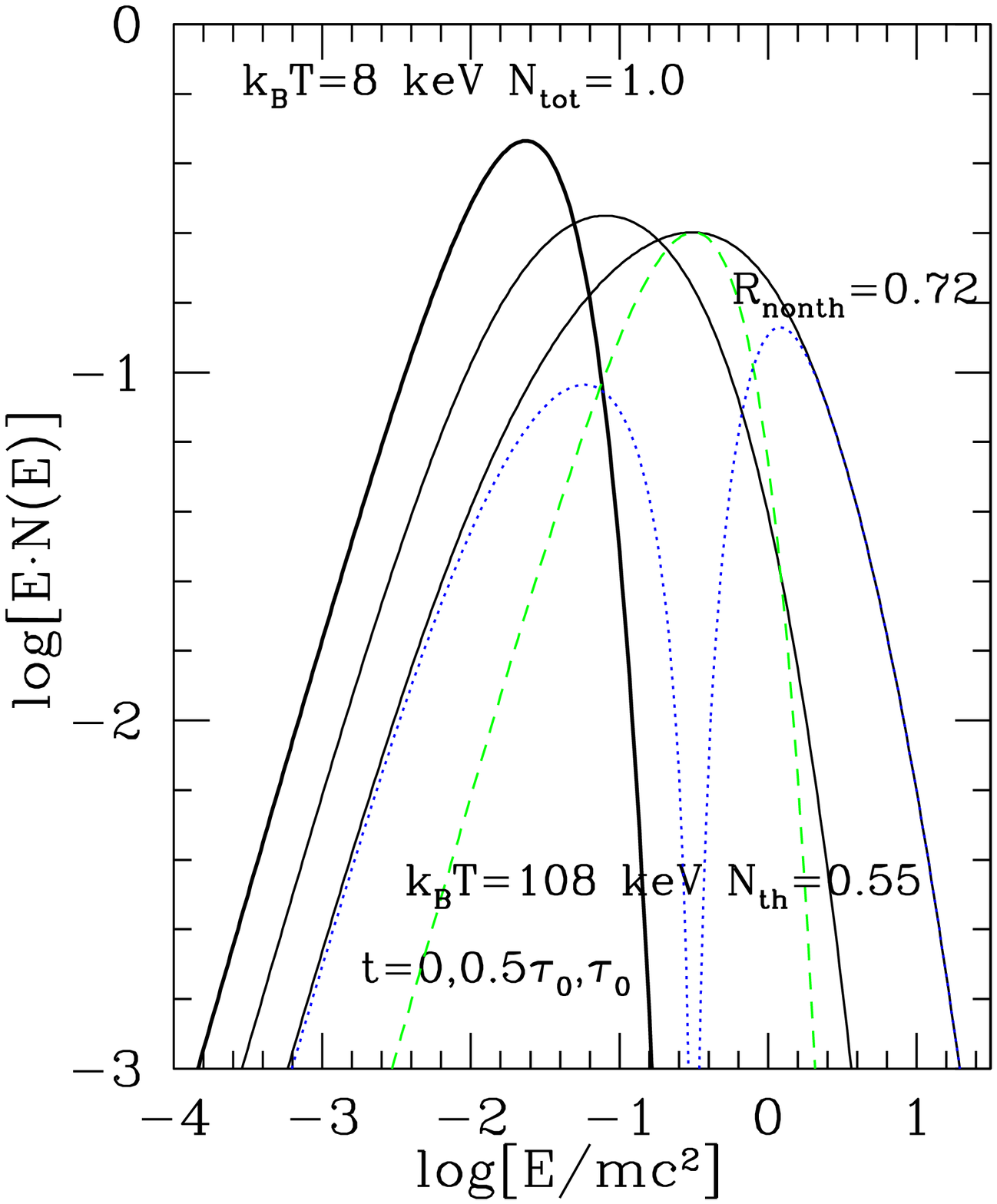}
\hspace{-1.4cm}
\includegraphics[height=6.2cm]{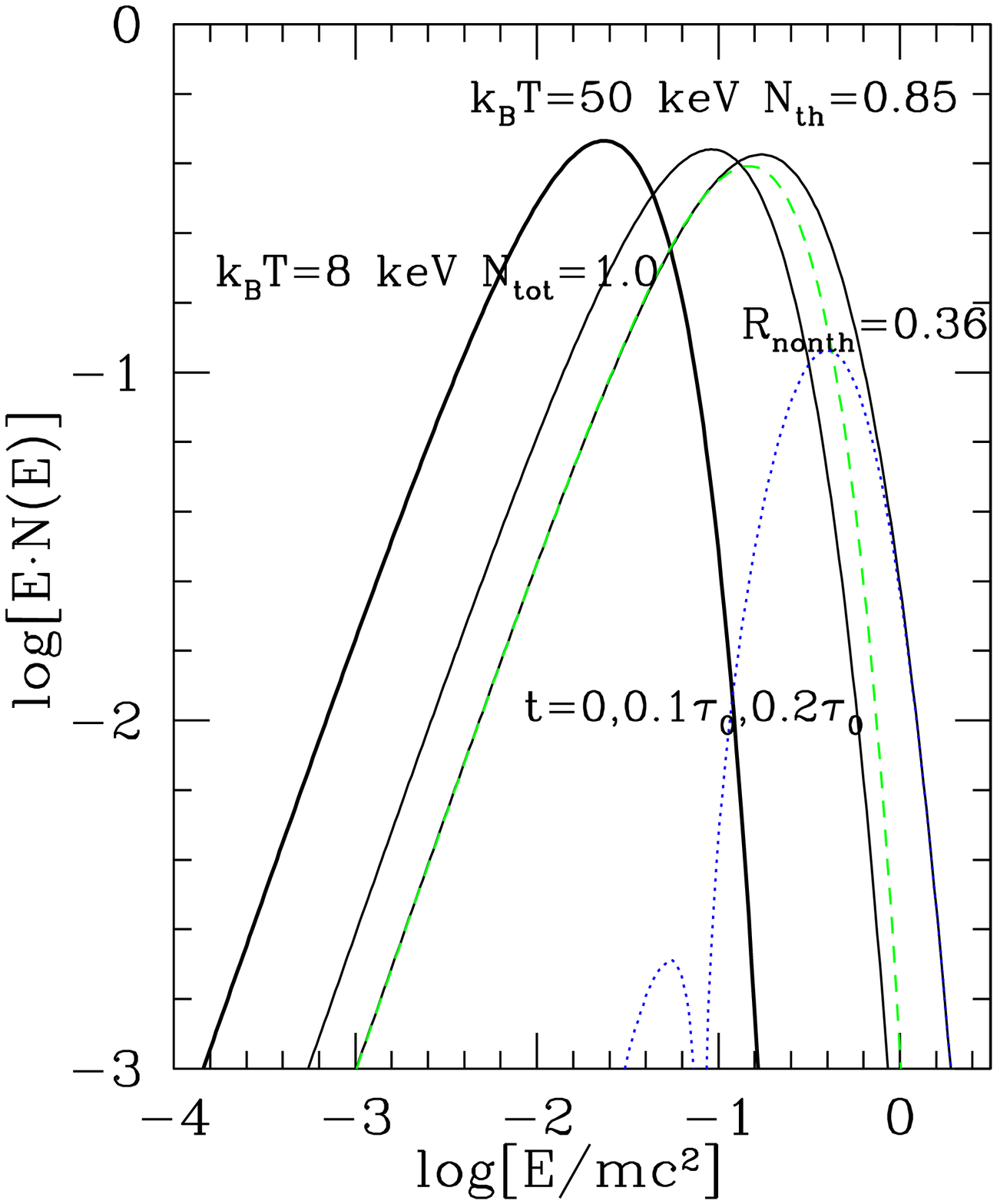}
\hspace{-1.4cm}
\end{center}
\caption{ Evolution with time of electron spectra in the presence
of a
constant level of turbulence that accelerates electrons according to equation
(\ref{acctime}) with $\tau_0/\tau_{\rm Coul}=0.013$, $E_c=0.2 (\sim 25$ keV) and
$q=1,0,-1$ respectively from left to right. The thermal fit and the nonthermal
residuals are shown by the
dashed green and dotted blue curves respectively.  In each figure we give
the initial and final values of the
temperature, the fraction of electrons in the thermal component $N_{\rm th}$,
and the ratio of energy of the nonthermal component to the thermal components
$R_{\rm nonth}$.}
\label{spectra1}
\end{figure}

\begin{figure}[htbp]
\begin{center}
\includegraphics[height=7cm]{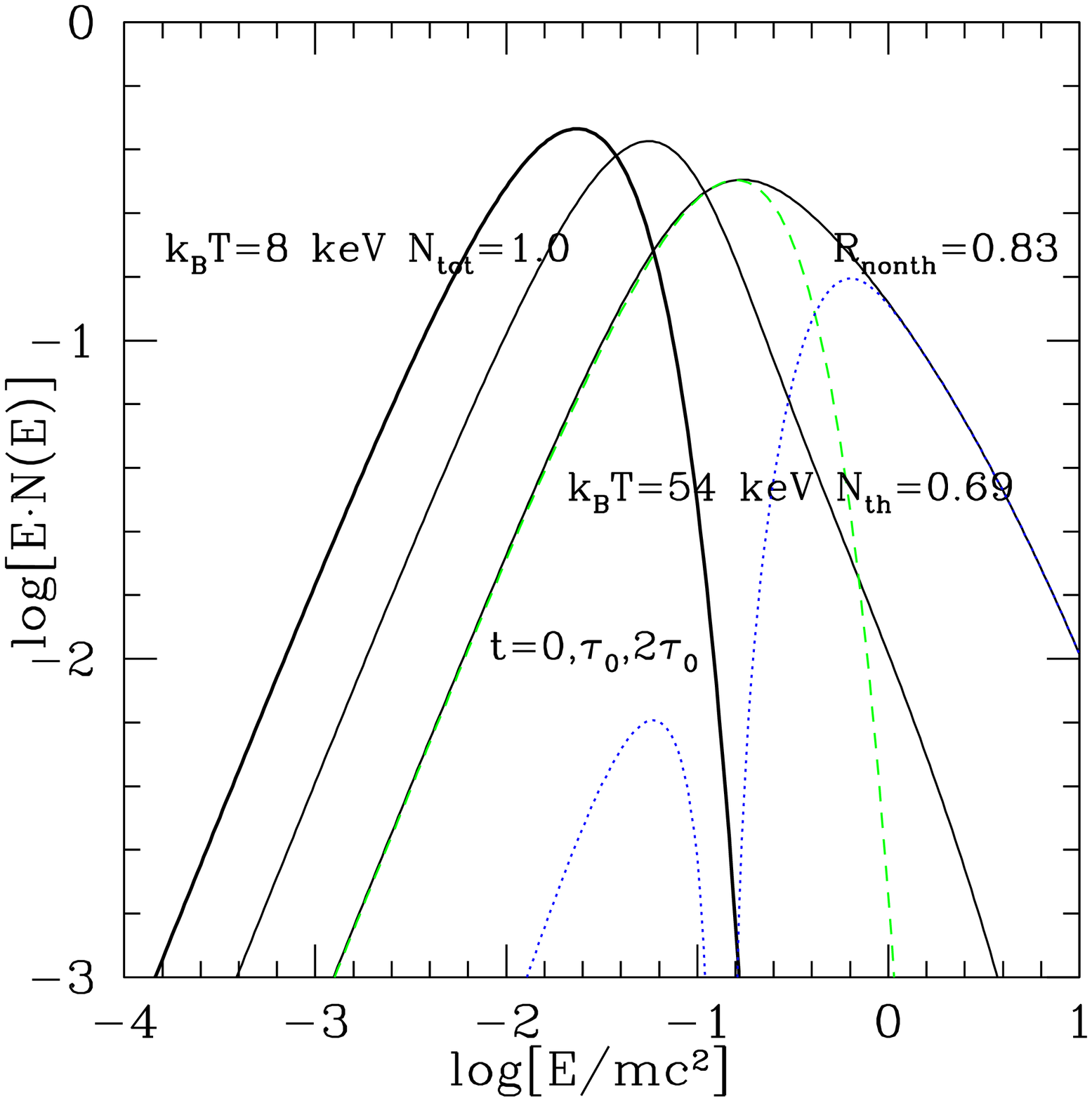}
\includegraphics[height=7cm]{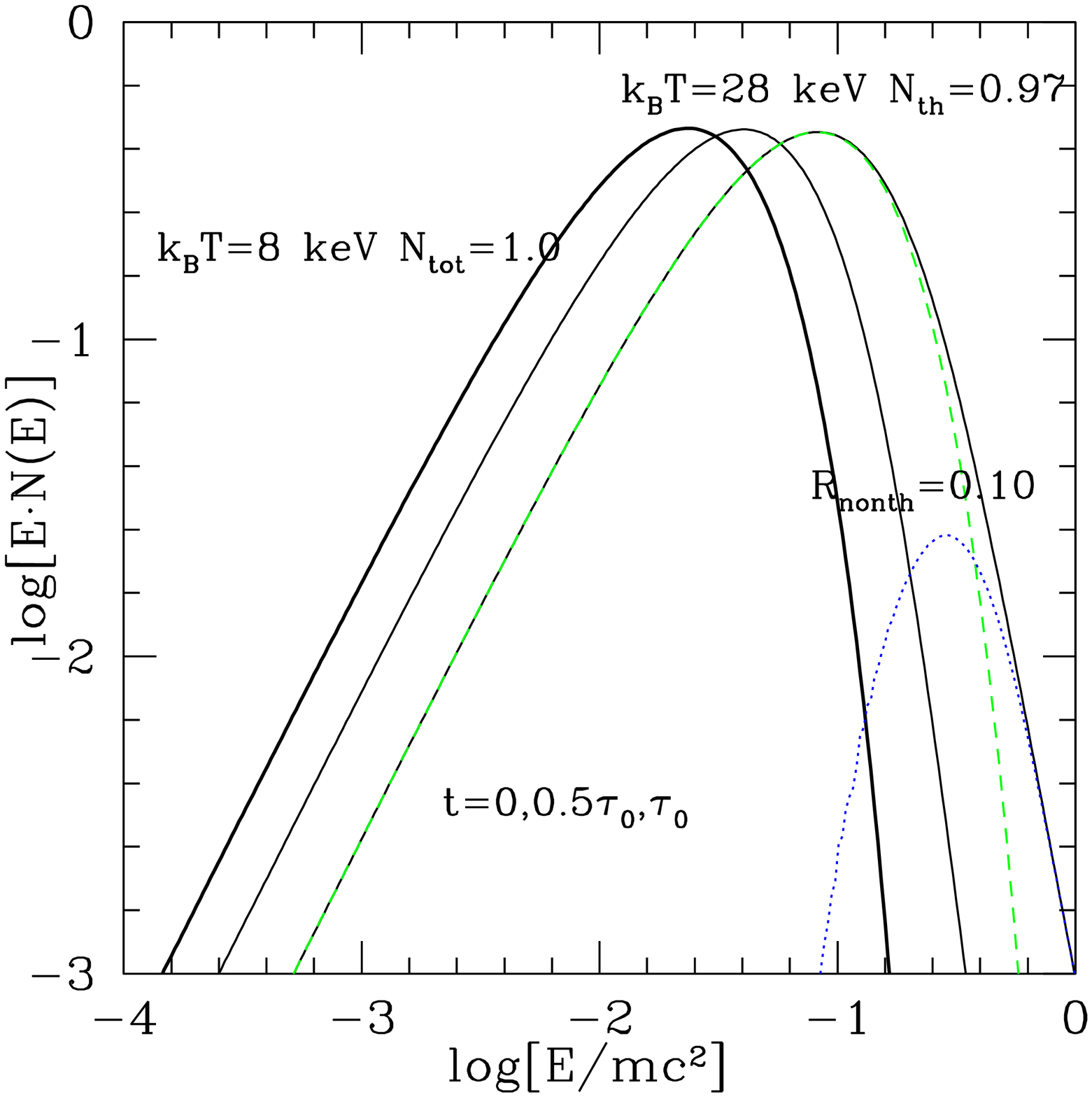}
\end{center}
\caption{ Same as the previous figure but with $q=1$ and 
$\tau_0/\tau_{\rm Coul}=0.18$ and $2.4$ respectively from left to right.}
\label{spectra2}
\end{figure}

Figures \ref{spectra1} and \ref{spectra2} show the evolution of initially Maxwellian distributions subject to various acceleration models.  Figure \ref{spectra1} shows the evolution for the three acceleration models
($q=-1,0,1; E_c=0.2$) shown in Figure \ref{timescales} (right panel) with the
smallest values of $\tau_0$ and Figure \ref{spectra2} shows the evolution for
the two other $\tau_0$'s but with $q=1$.  For each model we shown the spectrum at several evenly spaced time steps, beginning with the initial distribution.  In addition we plot a thermal fit to the final distribution and a nonthermal residual to this distribution. The general
feature of these results is that the turbulence causes both acceleration and
heating in the sense that the spectra at low energies resemble thermal
distribution
but have a substantial deviation from this quasi-thermal distribution at high
energies which can be fitted by a power law over a finite energy range (see
appendix).
Alternatively, one can fit the broad distribution by a multi-temperature model.
In most cases there is
no distinct nonthermal tail. In general the distributions are broad and
continuous, and  as
time progresses they become broader and shift to higher energies; the
temperature increases and the nonthermal `tail' becomes more prominent.
Comparing different values of the parameter $q$ which determines the low energy
behavior of the acceleration model we can see that for higher (lower) values of
$q$ the fraction of nonthermal particles is greater (smaller). For the highest
rate of turbulence $\tau_0=0.013\tau_{\rm Coul}\sim 10^5$ yr, as shown in Figure
\ref{spectra1}, the $q=1$ model
(with a reduced acceleration rate at low energies)  develops a large nonthermal
tail with a small amount of heating, while the $q=0$ model develops into a broad
distribution without a distinct nonthermal tail as well as heating up
significantly and the $q=-1$ model (with a higher acceleration rate at low
energies)mostly produces heating. For $\tau_0>\tau_{\rm Coul}$ there is
very little of a nonthermal tail and most of the
turbulent energy goes into heating. We fit all spectra to a best thermal
distribution and the
reminder is called the nonthermal part. The initial and final temperatures, the
fraction of particles in the quasi-thermal component $N_{\rm th}$, and the
ratio of nonthermal to thermal energies $R_{\rm nonth}$ are shown in Figures
\ref{spectra1} and \ref{spectra2}.
\clearpage

\clearpage

\begin{figure}[htbp]
\begin{center}
\includegraphics[height=6.2cm]{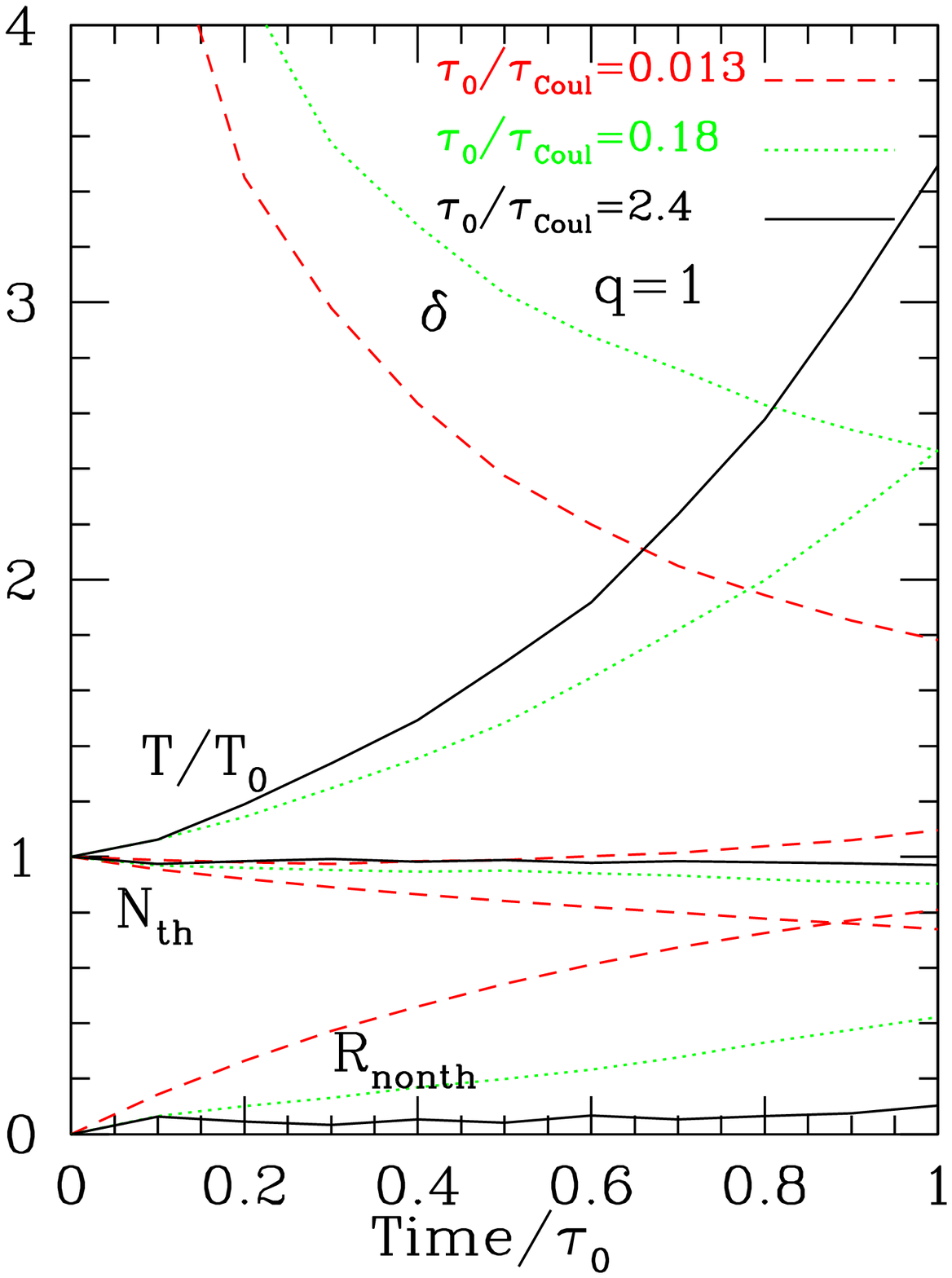}
\hspace{-1.4cm}
\includegraphics[height=6.2cm]{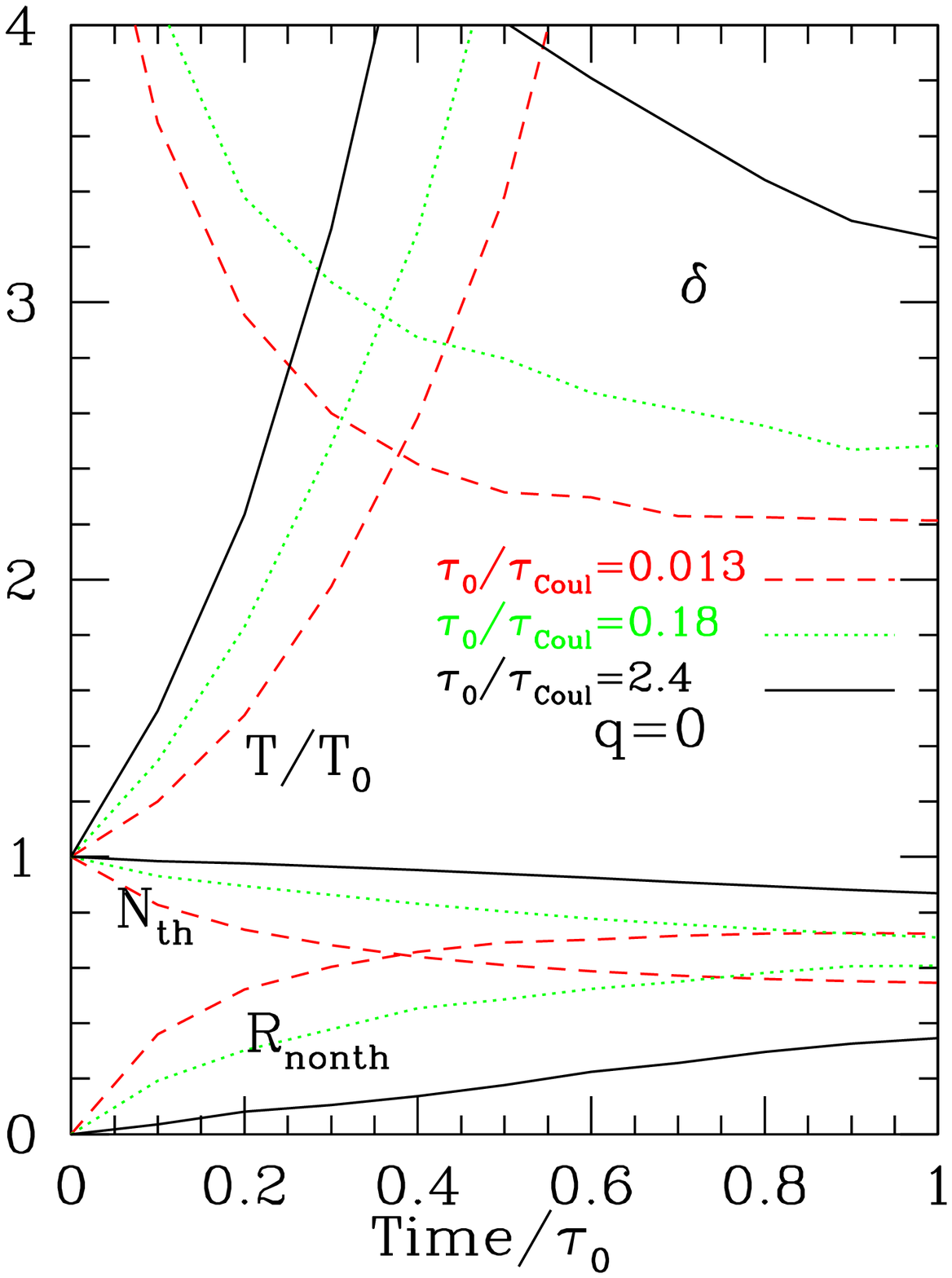}
\hspace{-1.4cm}
\includegraphics[height=6.2cm]{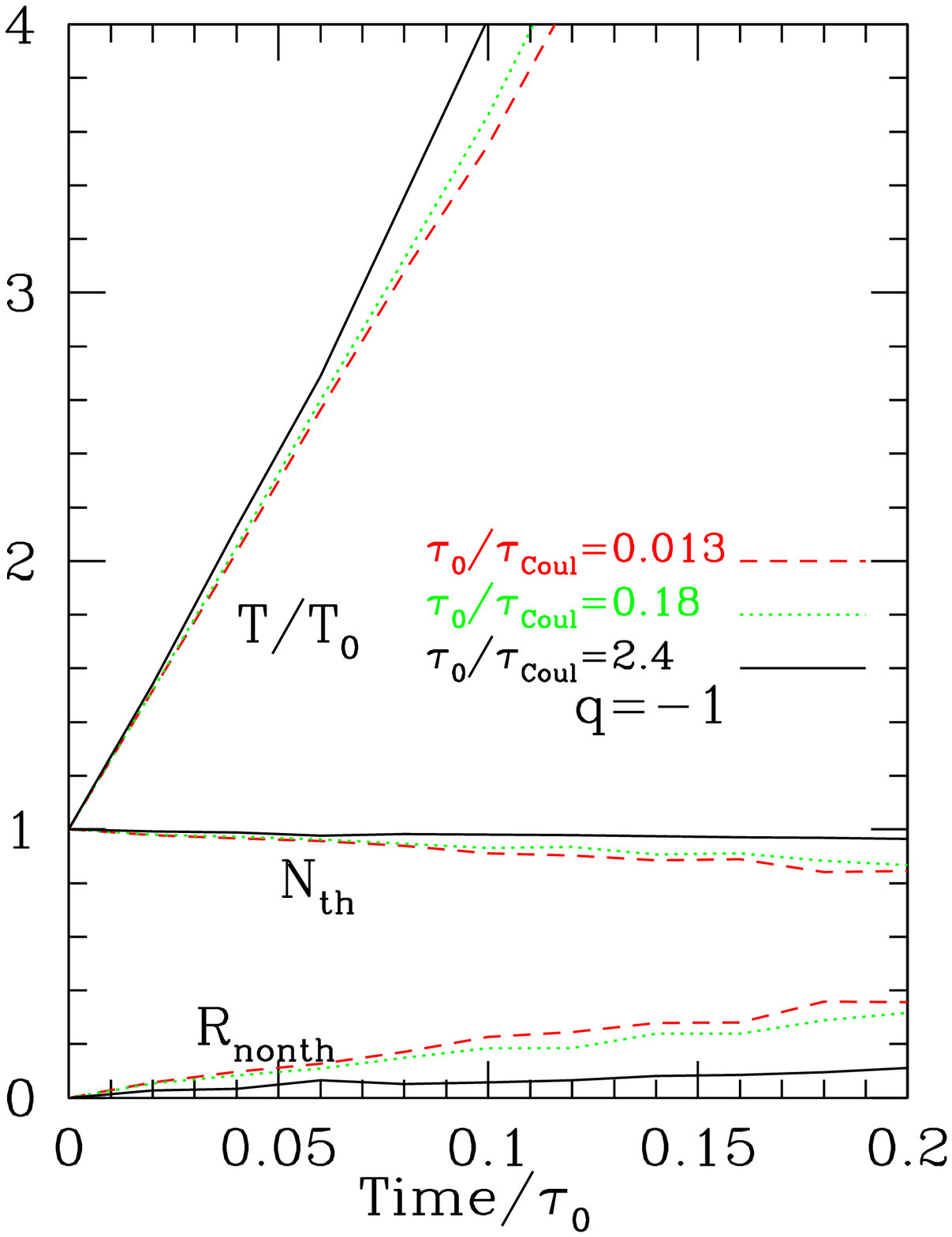}
\hspace{-1.4cm}
\end{center}
\caption{ Evolution with time (in units of $\tau_0$) of electron
spectral parameters, $T(t)/T_0, N_{\rm th}, R_{\rm
nonth}$ and the power-law index $\delta$ for indicated values of
$\tau_0/\tau_{\rm Coul}$ ratios and the parameter $q$ (and $E_c=0.2$). Note that
for models with the same value of $q$ at $t=\tau_0$ roughly the same amount of
energy has been input into the ICM. In all cases except for the one with $q=1$ and $\tau_0/\tau_{\rm Coul}=0.013$ the temperature increases by
more than a factor of 2. This factor is smaller at higher rates of acceleration.
In addition, high acceleration rates produce flatter nonthermal tails (smaller
$\delta$) and a larger fraction of nonthermal particles (smaller $N_{\rm th}$)
and energy ($R_{\rm nonth}$).}
\label{evolution}
\end{figure}

We calculate the above parameters as well as the index $\delta=-d\ln N(E)/d\ln
E$ for several time steps up to the time $t=\tau_0$. In Figure \ref{evolution}
we show the evolution  with time of these parameters for all nine of the acceleration models shown in the right panel of Figure \ref{timescales} grouped by the value of the parameter $q$.  From these figures we can see that in all cases except for the one with $q=1$ and $\tau_0/\tau_{\rm Coul}=0.013$ the temperature increases by more than a factor of 2 by $t=\tau_0$.  We can also see that faster acceleration rates lead to more pronounced nonthermal components with flatter tails (corresponding to a smaller $\delta$) more particles (corresponding to smaller $N_{th}$) and more energy (corresponding to higher $R_{nonth}$).  In addition it is evident that increasing the acceleration rate at low energies (by increasing $q$) leads to larger temperature increases.

As evident from these results in most cases there is a large rise in temperature
before a significant nonthermal tail is produced. Noteworthy among these results
is the case with a high acceleration rate and $q=1$ (which means that the
acceleartion rate increases with energy) where a promising spectrum to explain
the HXR observation is obtained. Unfortunately this spectrum appears after about
$<3\times 10^5$ years and at its rate of energization the electrons will achieve
relativistic temperatures and energies on timescales $>10^8$ yr.
\clearpage
\clearpage
\section{SUMMARY AND CONCLUSION}

The aim of this paper is to explore the possibility that the HXR excesses
observed in several clusters of galaxies could be produced by bremsstrahlung
emission by a population of `nonthermal' electrons. Since the observed HXR tails
can be fitted to a multi-temperature or one temperature plus a power-law model
the same will be true for the electron distribution. Based on cold-target
Coulomb loss rates in P01 we showed that these scenarios must be short lived
otherwise there will be extensive heating of the ICM gas. Here we carry out a
more detailed analysis of the problem including  the fact that for some low
energy electrons the cold target assumption is not accurate. We derive  exact
forms for the energy loss and diffusion coefficients for an arbitrary
distribution of electrons to be used in the particle kinetic equation for
determination of the spectral evolution of the electrons. We also devise
approximate recipes which
can be utilized more readily. We test our algorithm by evaluating the rate of
relaxation of an arbitrary distribution of particles into a Maxwellian one under
the influence of Coulomb collisions alone. The thermalization occurs within the
expected time.

We first evaluate the {\it survival time of nonthermal tails}, such as a
distinct high energy bump (Gaussian shape) or a power law, in a background
plasma with parameters appropriate for ICM conditions. We assume two conditions.
In one scenario, which we refer to as {\it test particle case}, it is assumed
that the energy of the tail is  insignificant, or its input is radiated or
conducted away, so that the background temperature remains constant. In a second
scenario  this assumption is relaxed and the evolution of the temperature is
explicitly determined. We compare the survival times of such tails with what one
would get in a cold-target scenario in P01. We find that the lifetimes of the
tails is increased by factors of two or three. {\it This reduces the severity of
the
difficulty in producing the HXR tails by nonthermal bremsstrahlung process
discussed in P01 but does not alleviate it completely.}

In P01 it was also claimed that because of the large Coulomb losses acceleration
of background thermal particles into a distinct nonthermal tail will  be
accompanied by a catastrophic heating, and  that long lived ($\sim 10^9$ yr)
nonthermal electrons must have energies greater than few 100's of MeV to avoid
this difficulty. Here we have carried detailed analysis of evolution of
nonthermal electron
spectra under the influence of some generic acceleration, or more correctly
energizing, process while suffering Coulomb (and all radiative) losses. The
results confirm the earlier claim;  in general an initial Maxwellian spectrum
evolves into a fairly broad spectrum without a distinct nonthermal tail. The
resultant spectra can be decomposed in many ways. In most cases the spectra are
dominated by a single temperature Maxwellian. Consequently we decompose the
spectra into a Maxwellian core (and determine its density and temperature) and
the remainder is lumped into a `nonthermal tail'. We refer to these components
as heated and accelerated electrons, respectively. Our results can be summarized
as follows:

\begin{enumerate}

\item At energizing rates smaller than the thermalization rate of the background
plasma there is very little acceleration. The primarily effect is heating of the
plasma at a rate equal to the energizing rate. Therefore, in order to avoid
excessive heating the
energizing process must have timescale must be comparable to the Hubble time or
be short lived and last less than the thermalization time which is $<10^5$
years.

\item A corollary of this is that in the steady-state situation, where the
energizing
rate is equal to the radiative loss rate, there will not be a nonthermal tail.
At the few keV temperatures the radiative loss is dominated by the
bremsstrahlung process which is very slow compared to the Coulomb scatterings.

\item At higher energizing rates a distinguishable nonthermal tail is developed
but this is again accompanied by an unacceptably high rate of heating. For
example, for $\tau_0\sim 5\times 10^6$ yr and $q=1$ about 10\% of
electrons end up in a nonthermal tail but the background temperature is
increased to $20$ keV from an initial value of 8 keV within five million years.

\item A well developed tail occurs only for the relatively fast energizing case
with $\tau_0\sim 10^{-2}\tau_{\rm Coul}$.
The heating rate for such cases is relatively slow; However, this phenomenon is
short lived and the vast majority of the particles will be accelerated to high
energies if this level of turbulence is maintained for times of order $\tau_{\rm
Coul}\sim 10^7$ yr. The only way to avoid a catastrophic heating will be if the
energy process lasts $\sim 10^5$ years.

\item But to explain the HXR observations by NTB would require episodic
energizing once every $\tau_{\rm therm}(E\sim 50 keV)\sim 10^6$ years before the
nonthermal tail is dissipated heating the plasma, so that at the end, on
average, this will amount to a hotter plasma and less of a nonthermal tail
similar to a case with slower acceleration.

\end{enumerate}

These results disagree with those
presented by Dogiel et al. (2007) where it is claimed that
nonthermal tails can be maintained for long periods and acceleration by
turbulence can produced power law tails without excessive heating. However,
their result
seem to be based on cold-target Coulomb loss rate. They also introduce a
energy diffusion term which does depend on a constant temperature and would be absent for a cold target case. Because this temperature is fixed, heating is not allowed. The crucial
fact that these coefficients depend on the exact shape of the particle spectra
is not included in their calculations. Our results also disagree with those of
Blasi (2000). This is somewhat puzzling because unlike the above paper Blasi
uses an algorithm similar to ours except that he calculates turbulence
coefficients based on an assumed spectrum of turbulence. In fact, we tried  to
reproduce his results using the given turbulence spectrum and the exact Coulomb
coefficients from NM98. To begin with, as noted in Wolfe \& Melia (2006), we
found that the entire spectrum became accelerated to higher energies on
timescales much shorter than $10^8$ yr unless a low energy cutoff in the
turbulence was introduced.  Following Wolfe and Melia's prescription and setting
the turbulence to zero for $\beta<0.5$ we still found much more heating than in
Blasi (2000). We found the temperature of the nonthermal component rose from
$7.5$ keV to $9.7$, $12.7$, and $44$ keV after $3$, $4$, and $5\times 10^8$ yr
respectively.  This is compared with a temperature of $8.2$ keV that Blasi's
results show after $5\times 10^8$ yr. This and our general results on the
acceleration agrees qualitatively with the results of Wolfe \& Melia (2006)
which also show that with such a turbulence model the electrons will be heated
to too high a temperature before $10^9$ yr to match observations.  However, we
should also note that these authors state that the Fokker-Planck and Coulomb
coefficient formulation based on NM98 that we use here may suffer from some
numerical problems.  For example, using NM98 coefficients they claim that an
initially Maxwellian distribution with $kT=0.1 m_ec^2$ subjected only to Coulomb
interactions changes by $15\%$ after 4 Spitzer times. We found however that we
were able to achieve less than $1\%$ deviation (defined here as $100\times \int
|N_{\rm th}(E)-N(E)|dE$) after $10$ Spitzer times.

Our analysis of this problem has been limited to consideration of energizing the
electrons. Eventually, on a  $\sim 2000$ times longer timescales  the electrons
will come into thermal equilibrium with protons and the estimated temperatures
will be reduced by a factor of two. We have not considered the possibility of
direct energizing of the proton by the same process that heats up the electrons.
The situation then becomes more complex and will depend on the relative rate of
energy input in electrons and protons.  However, we note that HXRs in the $>20$
keV range can also be produced via interactions between low energy thermal
electrons and nonthermal protons with energies greater than  40 MeV in a process
one may call {\it inverse bremsstrahlung}. In the rest frame of the protons the
electrons will have the requisite velocity to produce HXR photons. The Coulomb
loss rate of the nonthermal protons, being mainly due to their encounters with
the thermal protons, will be 43 times longer. This may increase the
bremsstrahlung yield by an undetermined factor which will depend on the details
of the electron and proton acceleration rates and energy dependence. Treating
this problem is beyond the scope of this paper where our main goal has been to
clarify the situation with the electrons as summarized above. In future works we
will consider this more complex problem of electron and proton acceleration.

\acknowledgements
We thank Siming Liu and Wei Liu for extensive discussion on the problem of
acceleration. This work was supported by the NASA grants NNG046A66G and
NNX07AG65G and by a  Stanford University VPUE grant.

\appendix
\section{TIME EVOLUTION OF TEMPERATURE}

When the Coulomb energy loss (gain) rate is given as a function of temperature
as in equation (\ref{hot}) an expression for the time evolution of this
temperature can be derived based on energy conservation. In the case where there
is no turbulence and the only energy exchange occurs through Coulomb
interactions the total energy of the system should remain constant.  The rate of
change in total energy of the system, $\frac{\partial {\cal E}_{\rm
tot}}{\partial
t}=\frac{\partial}{\partial t}\int_0^\infty N(E,t)EdE$, can be rewritten using
equation (\ref{KEQ}) along with the no-flux boundary condition (see Park \&
Petrosian 1995 eq. [3]) and the assumption that $D_{\rm Coul}(E)N(E)$ vanishes
at the
boundaries as
\begin{equation}
\label{change}
\frac{\partial {\cal E}_{\rm tot}}{\partial t}=-\int_0^\infty \dot{E}_{\rm
Coul}^{\rm
hot}N(E,t)dE.
\end{equation}
Now we can consider the Coulomb energy loss term to be a function of temperature
which is in turn a function of time, $\dot{E}_{\rm Coul}^{\rm hot}(E,T(t))$.  At
some initial time $t=t_0$ we assume that the temperature is set so that
$\frac{\partial {\cal E}_{\rm tot}}{\partial t} (t_0)=0$ (for example the
initial
distribution is Maxwellian and the energy loss term temperature is set to the
temperature of the distribution).  Therefore, in order to conserve energy at all
times we require that $\frac{\partial^2 {\cal E}_{\rm tot}}{\partial t^2}=0$. 
Based
on equation
(\ref{change}) this means that
\begin{equation}
\frac{\partial T}{\partial t}=-\frac{\int_0^\infty \dot{E}_{\rm Coul}^{\rm
hot}(E,T)\frac{\partial N}{\partial t}(E,t)dE}{\int_0^\infty \frac{\partial
\dot{E}_{\rm Coul}^{\rm hot}}{\partial T}(E,T)N(E,t)dE}.
\end{equation}
A new value of $T$ can be calculated at each time step and used for calculation
of the Coulomb rates (eqs. [\ref{DCoul}] and [\ref{eff}]) to be used in the
solution of equation (\ref{KEQ}) at the next time step.

\begin{figure}[htbp]
\begin{center}
\includegraphics[height=7cm]{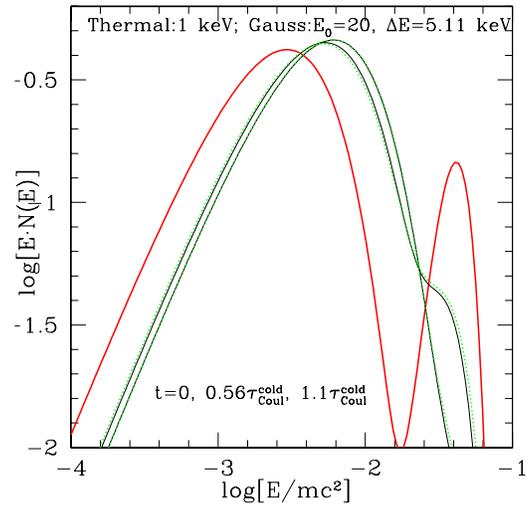}
\end{center}
\caption{A comparison of the two different methods for calculating time dependent Coulomb coefficients using the same conditions as the bottom left panel of Figure \ref{tests}.  The solid red line is the initial particle distribution.  The solid black lines were calculated according to the method described in the appendix which uses a time dependent temperature parameter.  The dotted green lines were calculated according to equations (\ref{general}) and (\ref{generaldiff}). }
\label{methodcomp}
\end{figure}

As long as the particle distribution is nearly Maxwellian, this method gives similar results to the more computationally intensive method of using equations (\ref{general}) and (\ref{generaldiff}) to calculate the coefficients at each time step.  Figure \ref{methodcomp} shows a comparison of these two methods using the same conditions as the bottom left panel of Figure \ref{tests}.

\clearpage
\clearpage

\section{FITTING METHODS}
The following provides details on the fitting methods used in \S 3.  Given a
particle distribution $N(E)$ the thermal component was determined by a two
parameter  fit.  Consider a Maxwellian distribution $N_{\rm th}(E)=A\sqrt{E}e^{-Em_ec^2/kT}$ where $A$ and $T$ are the two free parameters. 
For
$E
\ll \frac{kT}{m_ec^2}$ we have $\ln(N_{\rm th}(E))\approx
\frac{1}{2}\ln(E)+\ln(A)$.  Therefore $A$ was determined by finding
the intercept of the tangent line to the distribution in log-log space
approximately three orders of magnitude below the energy at which the
distribution peaks.  This insures that the particle distribution and the fitted
Maxwellian agree at low energies.  The value of $T$ was then determined by
finding the largest such value such that $N_{\rm th}(E)\leq N(E)$ across the
energy
range.  This gives the largest thermal component contained within the particle
distribution. Note that for the acceleration models with $q=0$ the distribution
was too broad to fit well to a Maxwellian at lower energies and $A$ and $T$
were instead determined by requiring the peaks of $N_{\rm th}(E)$ and $N(E)$ to
coincide. The total number of thermal particles is given by $N_{\rm th}=\frac{A
\sqrt{\pi}}{2}(kT/m_ec^2)^{3/2}$ and the total number of nonthermal particles is given
by $N_{\rm nonth}=1-N_{\rm th}$. The ratio of nonthermal energy to total energy
was
calculated as $R_{\rm nonth}=1-\frac{(3/2)kT}{{\cal E}_{\rm tot}}$ where
${\cal E}_{\rm tot}=\int_0^{\infty}N(E)EdE$.  The power law index was calculated
from the
nonthermal component $N_{\rm nonth}(E)=N(E)-N_{\rm th}(E)$ as 
$\delta=-\frac{d\ln
N_{\rm nonth}(E)}{d\ln E}$ at an an energy two orders of magnitude above the energy
where the nonthermal component peaks.

{}

\end{document}